\documentclass[showpacs,preprintnumbers,amsmath,reprint,aps,prl,superscriptaddress,twocolumn]{revtex4-2}

\usepackage{bm}
\usepackage{graphicx}
\usepackage{color}
\usepackage{amsmath}
\usepackage{hyperref}
\usepackage{epstopdf}
\usepackage{upgreek}
\usepackage{mathptmx, textcomp}
\usepackage[utf8]{inputenc}
\usepackage[T1]{fontenc}
\usepackage{array, multirow}
\usepackage[table]{xcolor}
\usepackage{booktabs}
\usepackage{longtable}
\usepackage{braket}
\usepackage{cancel}
\usepackage{soul}
\usepackage{lineno}

\usepackage[normalem]{ulem}

\newcommand{\commentOut}[1]{}

\begin{document}

\title{Steering reaction flux by coupling product channels}

\author{Dominik Dorer}
\affiliation{Institut f\"{u}r Quantenmaterie and Center for Integrated Quantum Science \\ and Technology IQ$^{ST}$, Universit\"{a}t Ulm, D-89069 Ulm, Germany}

\author{Shinsuke Haze}
\affiliation{Institut f\"{u}r Quantenmaterie and Center for Integrated Quantum Science \\ and Technology IQ$^{ST}$, Universit\"{a}t Ulm, D-89069 Ulm, Germany}
\affiliation{Center for Quantum Information and Quantum Biology, Osaka University, 1-2 Machikaneyama, Toyonaka, Osaka 560-0043, Japan}

\author{Jing-Lun Li}
\affiliation{Institut f\"{u}r Quantenmaterie and Center for Integrated Quantum Science \\ and Technology IQ$^{ST}$, Universit\"{a}t Ulm, D-89069 Ulm, Germany}

\author{Jos\'{e} P. D'Incao}
\affiliation{Institut f\"{u}r Quantenmaterie and Center for Integrated Quantum Science \\ and Technology IQ$^{ST}$, Universit\"{a}t Ulm, D-89069 Ulm, Germany}
\affiliation{JILA, NIST and Department of Physics, University of Colorado, Boulder, CO 80309-0440, USA}
\affiliation{Department of Physics, University of Massachusetts Boston, Boston, MA 02125, USA }

\author{Eberhard Tiemann}
\affiliation{Institut f\"ur Quantenoptik, Leibniz Universit\"at Hannover, 30167 Hannover, Germany}

\author{Paul S. Julienne}
\affiliation{Institut f\"{u}r Quantenmaterie and Center for Integrated Quantum Science \\ and Technology IQ$^{ST}$, Universit\"{a}t Ulm, D-89069 Ulm, Germany}
\affiliation{Joint Quantum Institute, University of Maryland and NIST, College Park, MD 20742, USA}
\author{Markus Dei{\ss}}
\affiliation{Institut f\"{u}r Quantenmaterie and Center for Integrated Quantum Science \\ and Technology IQ$^{ST}$, Universit\"{a}t Ulm, D-89069 Ulm, Germany}

\author{Johannes Hecker Denschlag} \email{johannes.denschlag@uni-ulm.de}
\affiliation{Institut f\"{u}r Quantenmaterie and Center for Integrated Quantum Science \\ and Technology IQ$^{ST}$, Universit\"{a}t Ulm, D-89069 Ulm, Germany}

\date{\today}
\begin{abstract}
We demonstrate a method for controlling the outcome of an ultracold chemical few-body reaction by redirecting a tunable fraction of reaction flux from one selected product channel to another one. 
 In the reaction, three ultracold atoms
 collide to form a diatomic molecule. This  product molecule can be produced in various internal states, characterizing the different product channels of the reaction. 
 Our scheme relies on the coupling between two such product channels at an avoided molecular energy level crossing in the presence of an external magnetic field. The degree of coupling can be set by the magnetic field strength and allows for a widely tunable flux control between the two channels. This scheme is quite general and also holds great promise for a large variety of chemical processes with diverse species, since molecular energy level crossings are ubiquitous in molecular systems and are often easily accessible by standard laboratory equipment.
\end{abstract}
\maketitle
Controlling chemical reactions at the quantum level has been a long-standing goal since the early days of quantum mechanics. 
Formally, a chemical reaction can be understood as reactants following one or several pathways along reaction coordinates (channels) in a quantum mechanical configuration space. Pathways can split up or merge along the way, e.g., due to local couplings between different channels, such that reaction dynamics are complicated, in general. 
For isolated, non-radiating systems, as typically realized in the dilute gas phase, the dynamics along pathways are coherent, and quantum interference effects are bound to occur. Generally speaking, the reactants approach each other from large internuclear distances, where they are asymptotically noninteracting. Via this entrance channel they enter the reaction zone at shorter distances where they form a reaction complex and the chemical dynamics take place. The products then leave the reaction zone via product (or exit) channels. Controlling the reaction flux into selected product channels is a central goal of the field.

Recent progress in ultracold, low-density atomic and molecular gases has led to a high level of control over several aspects of cold chemical reactions. This includes the preparation of reactants on the quantum level, tailoring their reaction environment (e.g. trap geometry, exposure to electric, magnetic or electromagnetic fields), as well as state-resolved detection  of the reaction products. After gaining tremendous control over cold chemical reactions involving two atoms \cite{Jones2006, Chin2010}, the field has been expanding towards more complex reactions involving larger numbers of atoms. Three-body resonances \cite{Greene2017,Ferlaino2011,dincao2018JPB} 
and
long-lived collision complexes  \cite{Mayle2013, Croft2017, Gersema2021,Gregory2020, Hu2019, Bause2021, Nichols2022,Yan2020} 
have been predicted and observed. In this context, the question arose
to what extent reaction outcomes of collision complexes are merely statistical and the final quantum state of the reactive system will be decohered and scrambled up \cite{Croft2014, Croft2017}.
Enhancement or suppression of the total rates of cold few-body chemical reactions via shielding of collisions
\cite{Anderegg2021, Lin2023, Schindewolf2022, Bigagli2023, Matsuda2020}
or via  various tunable scattering resonances \cite{Son2022, Yang2022, Park2023b, Chen2024, Cao2024} have been demonstrated. 
Besides globally turning a reaction on or off, it is also possible to coherently control the outcome of a reaction. Proposals in this direction can be found, e.g., in \cite{Shapiro2012, Yang2022b, Tscherbul2015, Hermsmeier2021}.
Recently, first schemes for controlling product state populations in ultracold gases have been experimentally demonstrated using quantum state-resolved preparation and detection of reactants and molecular products, respectively, e.g.  \cite{Wolf2017, Liu2021, Hu2021, Liu2024, Haze2022, Haze2025}.
These schemes exploited particular propensity rules and conservation laws dictating which exit channels are more likely to be populated for a given entrance channel. 

 \begin{figure}[t]
	\includegraphics[width=0.8\columnwidth]{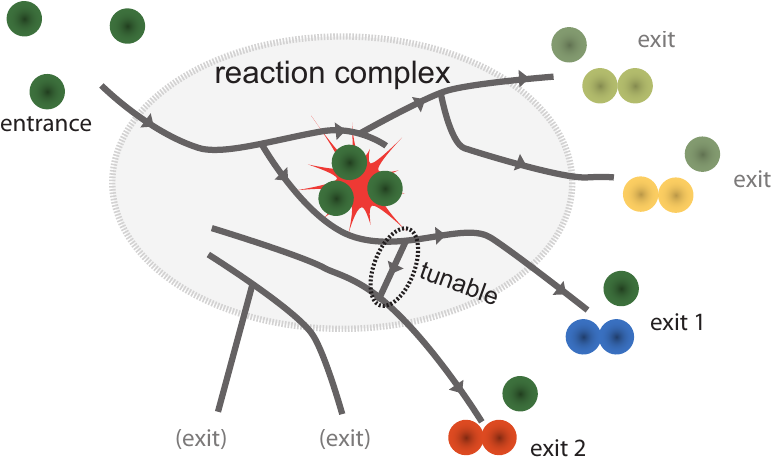}
	\caption{Schematic of a three-body recombination reaction, viewed in terms of coupled reaction channels. Three atoms on an entrance channel enter the reaction zone  where they form a reaction complex. Coupling of
     channels allows for the reaction to 
    propagate along various pathways. 
    Some couplings can be externally tunable (e.g. the coupling between the channels labeled exit 1 and exit 2, as marked by the dashed ellipse). 
    The particles leave the reaction zone on product/ exit channels, e.g., in terms of a diatomic molecule in a particular internal quantum state (as indicated by different colors) and a free atom. Exit channels denoted in parentheses are not connected to the entrance channel and therefore obtain no reaction flux.
    }
	\label{fig1intro}
\end{figure}

In this Letter, we demonstrate a new scheme which steers a few-body chemical reaction between precisely two selected product channels (see Fig. \ref{fig1intro}). The scheme can be viewed as a tunable beam splitter of a reaction pathway, in analogy to beam splitters in optics where a light beam can be coherently split up and recombined with another light beam. 
By near-resonantly coupling two exit channels,
we can divert a large fraction of the reaction flux from one particular exit channel  to the other one in a controlled manner. For this, we make use of a molecular level crossing which can be controlled via an external magnetic field.

 Our approach is quite general and flexible. First, avoided energy level crossings are ubiquitous in chemical systems, including polyatomic species. Second, besides magnetic fields other control knobs can be introduced, such as optical, microwave, or radiofrequency fields. 
 It should therefore be possible to couple almost arbitrary pairs of quantum states at any reaction coordinate and bound state energy. In addition, we expect our general scheme to be quite robust regarding, e.g., variations of the collision energy.
 
Our scheme is complementary to another recent reaction control scheme  \cite{Haze2025} which uses a Feshbach resonance. Feshbach control affects the reaction quite differently as it acts in the initial phase of a reaction, controlling reaction flux between whole spin families of product states but not exclusively between two particular exit states. Furthermore, Feshbach control is very sensitive to the collision energy of the reactants.

In our experiment we investigate three-body recombination (TBR) of ultracold $^{87}$Rb, where three atoms collide and two combine to form a weakly-bound Rb$_2$ molecule. The molecule and the third atom carry away the released binding energy in terms of relative motion.
TBR is an intrinsically complex few-body process. However, as shown in our previous work, for the given species and parameter regime,  TBR can to some extent be understood in terms of two-body physics. In brief, as the atom pair forming the molecule undergoes a two-body collision, it is perturbed by the collision with the third atom. The third atom interacts with the pair only via mechanical forces, so that the spin state of the pair is conserved. This simplification offers some valuable physical insight into our control scheme \cite{Haze2023,Haze2022}. 

The colliding atom pair in scattering state  $|S \uparrow \rangle $ (see Table \ref{tab:table1}) can relax to a molecular bound state, to the extent that the pair's original spin state $|\uparrow \rangle$ overlaps with the
molecular spin state. This gives rise to a spin-conservation propensity rule as previously observed in \cite{Haze2022, Wolf2019, Wolf2017}. 
By tuning the spin state $ | \uparrow \rangle$ amplitude in the 
molecular state, we can control the reaction flux into this state. 
This can be accomplished by using a magnetically tunable avoided crossing of two neighboring molecular levels and 
an externally applied magnetic field, see Fig. \ref{fig2}(a). 
The avoided crossing  consists of two bare hyperfine molecular levels
$|B\uparrow \rangle $ and  $|B\downarrow  \rangle $,
 with orthogonal spin states $|\uparrow \rangle$ and $|\downarrow \rangle$ (see purple and green dashed lines) with different magnetic moments.
The levels are coupled via spin exchange interaction which mixes them coherently and leads to level repulsion. As a result of the avoided crossing, we obtain an upper and a lower molecular level branch denoted by $|U\rangle$ and $|L\rangle$. The spin compositions of the two branches vary with magnetic field.  Figure  \ref{fig2} (b) shows the
$|\uparrow \rangle$ content in terms of the integral
\begin{equation}
	% P_{olp}(U/L) =
     \left| \braket{ \uparrow   | U/L } \right|^2 := 
    \int \left|\braket{\uparrow | \Psi_{U/L}\left(\vec{r}\right)}\right|^2 d^3\vec{r}
    \label{eq:1}
\end{equation}
for the two branches $|U\rangle$ or $|L\rangle$ of the avoided crossing as a function of the magnetic field. Here, $\Psi_{U/L}(\vec{r}) = \langle \vec{r}| U/L\rangle $ is the normalized molecular wave function for $| U \rangle$,  $| L\rangle$, respectively. As a consequence, we expect the reaction flux into each of the two levels to follow the respective integral from Eq. (\ref{eq:1}) as a function of magnetic field. 
In  Fig.  \ref{fig2}(a) the reaction fluxes into the two branches are qualitatively indicated by the widths of the arrows for three magnetic fields.

\begin{figure}[t]
	\includegraphics[width=\columnwidth]{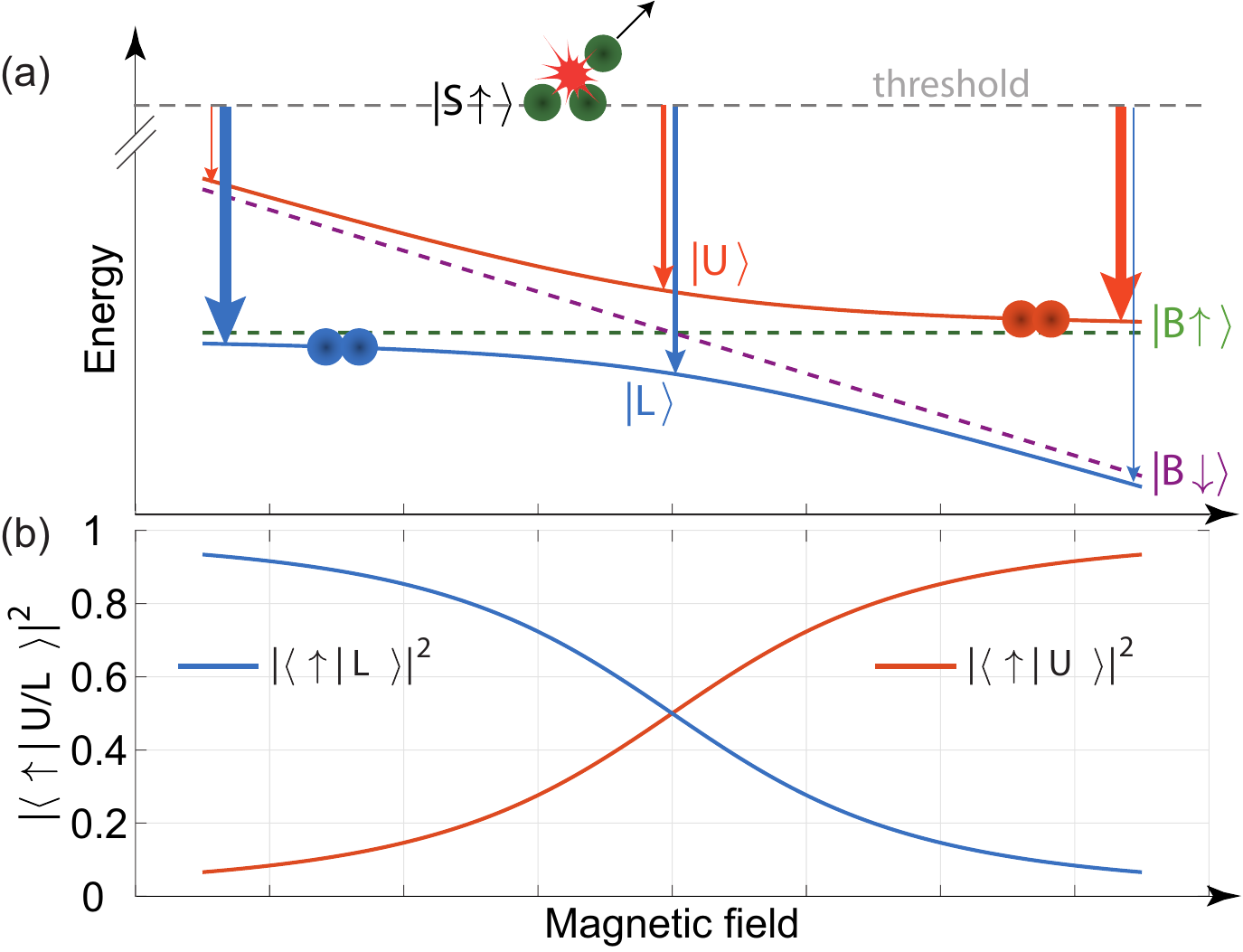}
	\caption{
    Schematics of the proposed reaction control scheme. 
    (a) The two bare states  $|B\uparrow \rangle $
    and $|B\downarrow \rangle $ are coupled and form an avoided crossing as a function of the magnetic field. This results in a lower and an upper branch, $| L \rangle $ and 
     $| U \rangle $, respectively. These two branches correspond to two exit channels of the three-body recombination reaction starting from the scattering state $|S\uparrow \rangle $. By tuning the external magnetic field, the reaction flux can be continuously steered between $| L \rangle $ and  $| U \rangle $. The widths of the vertical arrows illustrate this behavior in a qualitative way.
     (b) Calculated spin contents $ \left| \braket{  \uparrow | U} \right|^2$ and $ \left| \braket{    \uparrow  | L} \right|^2$  which correspond to the 
      $| \uparrow \rangle$ components of $| U \rangle $ and  $| L \rangle $. }
	\label{fig2}
\end{figure}

For our experiments we prepare an ($800\pm 50$) nK-cold gas of $^{87}$Rb atoms which are spin-polarized in the hyperfine state $f=1, m_f=-1$ of the electronic ground state $5S_{1/2}$. The sample of typically $(3.4\pm 0.1)\times 10^{6}$ atoms is trapped in a far-detuned optical dipole trap at 1065\,nm and has a peak particle density of $(8\pm 1)\times 10^{13}\:\mathrm{cm^{-3}}$. In the atom cloud, three-body recombination spontaneously occurs and preferentially produces weakly-bound molecules within the coupled electronic states $X^1\Sigma_g^+$ and $ a^3\Sigma_u^+$  \cite{Wolf2017,Haze2023}. More than 90\% of these molecules can be found in a range of vibrational (v =-1, $\cdots$, -6 ) and rotational ($L_R = 0, \cdots, 8$) quantum numbers \cite{Wolf2017, Haze2022, Haze2023, Haerter2013b} where $L_R$ corresponds to the
total angular momentum of the molecule, excluding electronic and nuclear spins.
Furthermore, we note that a negative vibrational quantum number means that we count
the vibrational levels downwards from the respective atomic threshold, so that  v=-1 represents the most weakly bound state for the considered atom pairs.

   \begin{table}[] %[h!]
   	\begin{center}
   		\begin{tabular}{| l|c|c|c|c|c|c|}
   			\hline
   			State & $F$ & $m_F$ & $f_a$ &  $f_b $ &  v  & $L_R $ 	\\
   			\hline 	\hline
            $|\uparrow \rangle $  & 2 & -2 & 1  & 1 &  - & -   \\
            $|\downarrow \rangle $  & 2 & -2& 2  & 2 & - & -   \\
   			$|S\uparrow \rangle $  & 2 & -2 & 1  & 1 &  - & 0   \\
   			$|B\downarrow \rangle $  & 2 & -2& 2  & 2 & -5 & 6   \\
   			$|B\uparrow \rangle $ & 2 & -2& 1  & 1 & -3 & 6  \\
   			$|R\uparrow \rangle $  & 2 & -2 & 1 & 1 & -4 & 4  \\
   			\hline
   		\end{tabular}
   			\caption{Quantum numbers of the spin states $|\uparrow \rangle $ and $|\downarrow \rangle $, the
            scattering state $|S\uparrow \rangle $, the bare molecular levels $|B\uparrow \rangle $ and  $|B\downarrow \rangle $,
            and the reference level $|R\uparrow \rangle $. Here, $F$ is the total angular momentum quantum number of the atom pair or molecule (excluding rotation) and $f_{a,b}$ are the total angular momentum quantum numbers of the two atoms (a,b) forming the molecule, respectively. }
   		\label{tab:table1}
   	\end{center}
   \end{table}

The bare states $|B\uparrow \rangle $ and  $|B\downarrow  \rangle $
are two hyperfine levels located within the described range of molecular bound states. 
Their quantum numbers  are given in Table \ref{tab:table1}.
Figure \ref{fig3} (a) shows the avoided crossing which is located at around 110 G. The symbols are experimental data, where we measured the binding energies of the two branches relative to the atomic $| S \uparrow \rangle$ threshold for each magnetic field.
The solid lines are coupled-channel calculations.
The experimental data are from high-resolution molecular spectroscopy where we state-selectively detect molecules in these levels, following their formation in three-body recombination. For this molecular spectroscopy, we use resonance-enhanced multiphoton ionization (REMPI) where one photon of a cw laser at a wavelength of $598\:\textrm{nm}$ resonantly excites the intermediate level  $(2)^1\Sigma_u^+$, v$'$ = 36, $J'$ = 5, where $J'$ denotes the 
quantum number of the total angular momentum of the molecule, excluding nuclear spins.
This transition step is strongly saturated in most cases.
Afterwards, a second identical photon ionizes the molecule. The produced ion is subsequently detected. From the resonant laser frequencies and an additionally measured photoassociation line to the same intermediate level the binding energies can be determined \cite{Wolf2017}.

\begin{figure}[]
	\includegraphics[width=0.85\columnwidth]{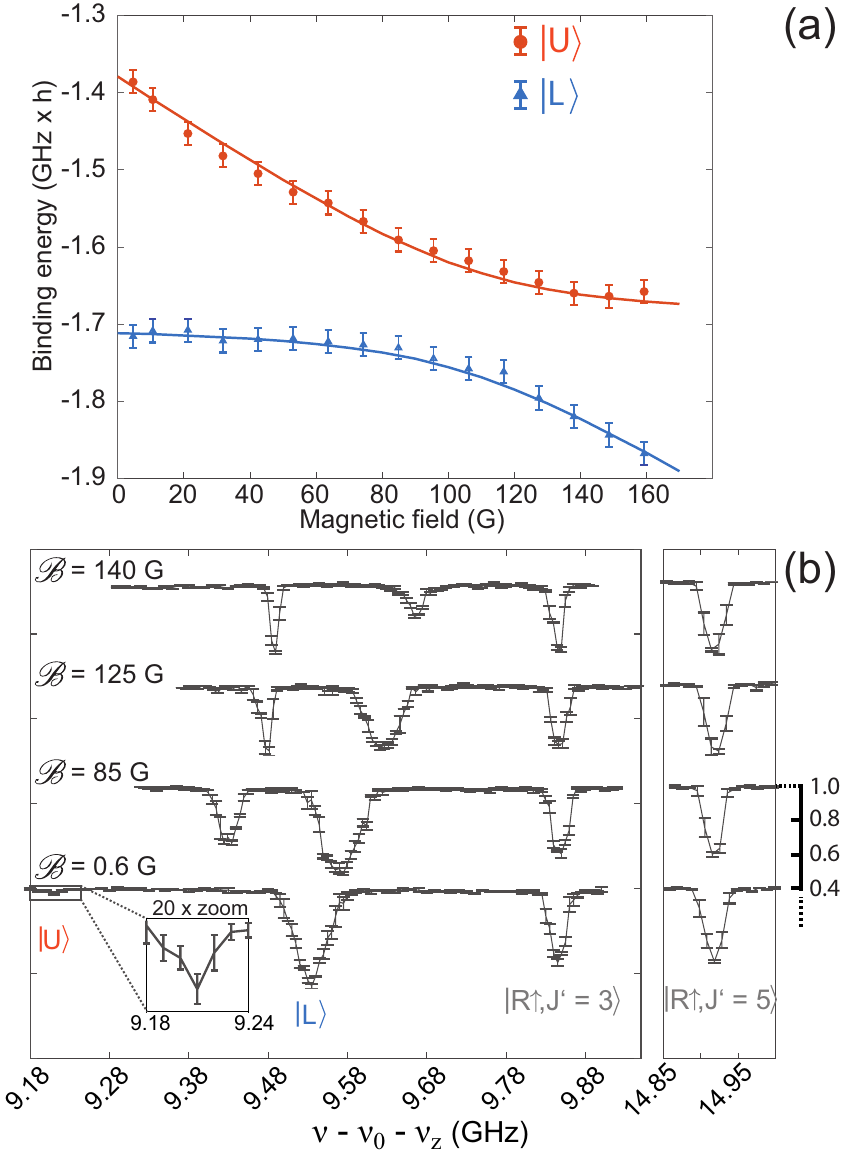}
	\caption{
		(a) Avoided crossing of the bare states $| B \uparrow \rangle$
		and  $| B \downarrow \rangle$, giving rise to an upper branch level $|U\rangle$ (red) and a lower branch level $|L\rangle$ (blue). The solid lines are calculated binding energies from a coupled-channel calculation. The data points are binding energies extracted from REMPI spectroscopy, where resonance positions were measured in small frequency steps with three repetitions of the experiment per frequency setting. Here, $h$ is Planck's constant. The error bars mainly reflect the typical uncertainty of the laser frequency and have been extracted from lorentzian fits of the resonance positions (95\% confidence level).
        (b) REMPI spectra of the avoided crossing as a function of the laser frequency $\nu$ for four different magnetic fields $\mathcal{B}$. The REMPI spectra show the normalized atom number with a scale as indicated for the spectrum at the magnetic field of 85 G. Each data point represents the mean value of five repetitions of the experiment and the error bars correspond to one standard deviation.
        Four spectra corresponding to different magnetic fields are presented. The 
        leftmost dip and the
        adjacent dip correspond to $|U\rangle$ and $|L\rangle$, respectively.The two dips on the right correspond to a reference level, $| R \uparrow \rangle$, detected via the intermediate levels $J'=3$ and $J'=5$ \cite{Rescaling}, respectively. The offset $\nu_0$ = 500.974420(10)\,THz is the photoassociation transition frequency towards the intermediate state  $(2)^1\Sigma_u^+$, v$'$ = 36, $J'$ = 1 at zero magnetic field. The offset $\nu_Z$ is the Zeeman shift of the reference level which corresponds to the Zeeman shift of the atom pair $(f=1,m_f=-1)+(f=1,m_f=-1)$.
        }
	\label{fig3}
\end{figure}
    
The REMPI photons stem from a $(120\pm 2)\,$mW laser beam with a diameter of approximately $(100\pm 10)\,\mu$m in the detection zone. The laser is frequency-stabilized to a wavelength meter and we observe a shot-to-shot frequency uncertainty of about 5 MHz (standard deviation). The detection of the produced ions works as follows. Immediately after production, the ions are trapped in a linear Paul trap which is centered on the atomic cloud. The trapped ions elastically collide with the atoms, leading to atomic loss which is measured, for details, see e.g. Ref. \cite{Haze2023}. 
By scanning the REMPI laser frequency $\nu$ over the optical transitions of the molecular levels  we obtain REMPI spectra. Figure \ref{fig3}(b) shows such spectra for four magnetic fields \cite{Bfield}. 
Each spectrum exhibits four resonance dips which are molecular signals. The leftmost dip of each spectrum corresponds to the upper branch $|U \rangle$ of the avoided crossing. The next dip to the right corresponds to the lower branch $|L\rangle$. The two other signals stem from the molecular state $| R \uparrow \rangle$ with a binding energy of $\approx 7.1\,\textrm{GHz}\times h$, which was used as a reference, see Table \ref{tab:table1}. This state is detected via the intermediate level $J'=3$ (second to rightmost dip) and $J'=5$ (rightmost dip), respectively. All these measurements are carried out with the same REMPI laser intensities and pulse lengths \cite{Rescaling}.
In general, the deeper the dip, the higher is the molecular production rate. 
This relation, however, is nonlinear and is generally subject to saturation effects.
The spectra indicate that the molecular production rate grows with increasing magnetic field for the upper branch, but diminishes for the lower branch. In contrast, the signals for the reference level $| R  \uparrow \rangle$ remain rather constant both for the detection via the $J'=5$ and $J'=3$ REMPI intermediate state, respectively.  
This suggests that for the reference level $| R  \uparrow \rangle$ 
(as well as for the bare states $| B  \uparrow (\downarrow) \rangle$ ) the reaction flux and the REMPI efficiencies are quite independent of the magnetic field.

For the  quantitative analysis, we have carried out another set of experiments, where we measured the ion production rate for the individual molecular states at the resonant REMPI laser frequencies [center positions of the dips in Fig. \ref{fig3}(b)]. Again, the produced ions are collected in the Paul trap. They are detected, however, by immersing them into a freshly prepared low-density ($(1.0 \pm 0.1)\times 10^{13}\mathrm{cm^{-3}}$) atomic cloud, where three-body recombination is decreased so much that atom loss  can be attributed mainly  to elastic atom-ion collisions. We obtain the number of ions by comparison to atom loss calibration measurements carried out with laser cooled Ba$^+$ ions which can be counted using fluorescence detection, as described in \cite{Haze2023}. From the ion numbers we derive ion production rates, which are directly proportional to the molecule formation rates to the extent that the REMPI efficiency is constant.

 \begin{figure}[t]
	\includegraphics[width=0.85\columnwidth]{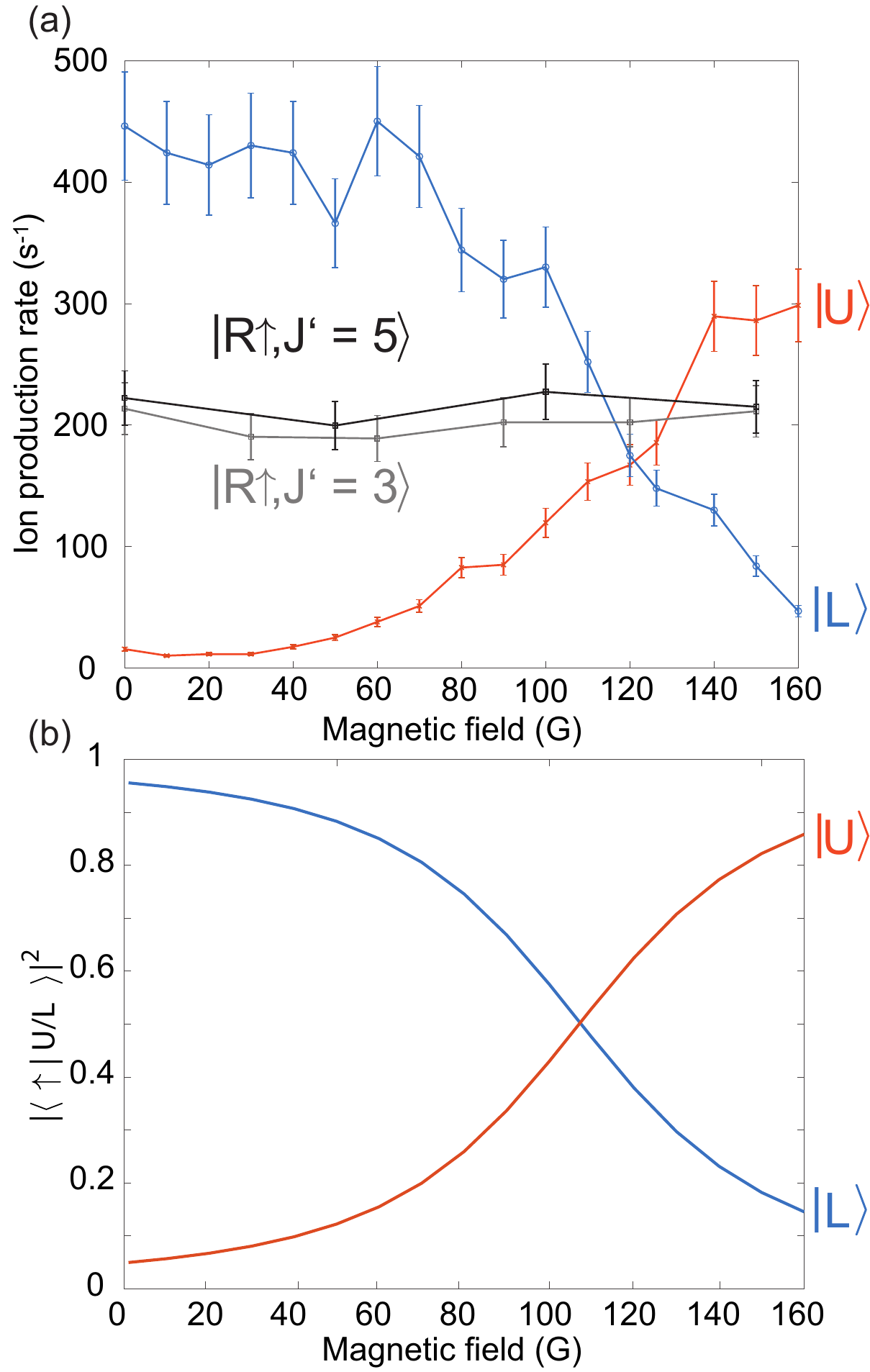}
	\caption{(a) Measured ion production rates for the three product states  $|U\rangle$, $|L\rangle$ and $|R \uparrow \rangle$. The reference state $|R \uparrow \rangle$ is detected via both the intermediate state $J'=3$ and  $J'=5$ \cite{Rescaling}.The error bars are the standard deviations. The experiment was repeated 30 times for each data point.
    (b) Calculated molecular $| \uparrow \rangle$ content
    for the upper and lower molecular state, $\left| \braket{  \uparrow  | U/L  } \right|^2$, as in Eq. (\ref{eq:1}). For the calculation we  used a two-body coupled channel Schrödinger equation with full molecular potentials.}
	\label{fig4}
\end{figure}

Figure  \ref{fig4}(a) shows measured ion production rates, as a function of the  magnetic field. Here, for each magnetic field we adjusted the REMPI laser pulse length in order to typically create less than five ions for a molecular state since our ion counting scheme works best for low numbers of ions, see Ref. \cite{Haze2023}.
Within the measured magnetic field range, the  production rate for the upper branch grows by a factor of 19 $\pm$ 3 with increasing magnetic field. At the same time, the production rate decreases correspondingly for the lower branch. 
The rates for the reference level  $| R \uparrow \rangle$ via $J' = 3$ and $J' = 5$  remain constant within a variation of 10 \%, in agreement with our more qualitative observations in Fig. \ref{fig3}. These results highlight the large tuning capability of our scheme for diverting  the reaction flux between two product channels. When adding the signals for the states 
$| U \rangle$ and 
$| L \rangle$ it becomes apparent that the total rate decreases for fields larger than $\approx$120 G such that
$\approx$20\% is missing at 160 G.
Judging from our theoretical calculations
(which will be discussed in the following) and from 
the constant signal of $| R   \uparrow \rangle$, such a  decrease in the molecular production rate is not expected. The missing signal can, however, be explained by a 
corresponding decrease of REMPI efficiency for the 
state $| U  \rangle$ due to a quantum interference effect which lowers the transition matrix element from $| U   \rangle$ towards the intermediate state (see Supplemental Material \cite{Supplemental}). For the other magnetic fields, as well as for the states $| L   \rangle$  and $| R   \uparrow \rangle$
at all magnetic fields the first REMPI step is strongly saturated.

It is instructive to compare the measured molecule production rates of Fig. \ref{fig4}(a)  with the size of the 
$|\uparrow \rangle$ spin component of the product molecule,
as, according to our simplistic reaction model,
the reaction flux into a channel should be directly proportional to the spin state overlap of the entrance and exit channel. Using Eq.\eqref{eq:1},
we calculate the spin contents
$\left| \braket{  \uparrow  | U/L  } \right|^2$
for the two molecular branches.
Our results are obtained from numerical two-body coupled channel calculations using accurate Born-Oppenheimer potentials, and are shown in Fig. \ref{fig4}(b). 
Indeed, the spin component curves in (b) resemble the formation rate curves quite well. The position at which the molecular production rates cross is shifted by about 8\,G compared to the position where the molecular $| \uparrow \rangle$ spin content of both channels is equal, but this could be an artifact of the lowered REMPI efficiency for the state  $| U \rangle$  at higher fields. Therefore, the overall agreement suggests that the simplistic model can indeed be used to explain and design the reaction control, at least to first order. This is quite remarkable, since  the simplistic reaction model is based only on two-body physics. 

 The detailed investigation, however, reveals a discrepancy concerning the strict linearity between the production rate of a molecular state and its spin content. At low magnetic fields the ratio of the experimentally observed molecular production rates in both channels (28.4 $\pm$ 4) in Fig. \ref{fig4}(a)
is by a factor of 1.5 different from the spin content ratio (19.5) in Fig. \ref{fig4}(b).  This discrepancy seems to point to additional physics beyond the simplistic model which relies on pure two-body physics. To gain additional insights, we have conducted numerical three-body calculations for the production rates of the individual molecular states. The model that we use for these calculations (see Supplemental Material \cite{Supplemental}) is already quite sophisticated as it includes all relevant spin degrees of freedom and uses realistic pairwise two-body interactions down to short distances of 40$a_0$. 
While the calculations again qualitatively agree with the swapping of the production rates between the two channels at the avoided crossing and therefore confirm our general control scheme,
quantitatively there are deviations from the results of  both the simplistic model as well as experiments (see Supplemental Material \cite{Supplemental}). This could point towards effects of pure three-body interactions which are not captured by pairwise two-body interactions as used for our calculations.
Furthermore, discrepancies can arise from approximations that have to be made for the calculations such as the truncation of molecular potentials to a low number of bound states.
Future experimental and theoretical work should help to identify the specific reasons for the quantitative discrepancies and will lead the way to optimized control schemes.

In conclusion, we have demonstrated a scheme to divert the reaction flux between two product exit channels in a controlled fashion by coupling these product channels. Our scheme can be considered a local beam splitter for the reaction pathway. Here, we have already demonstrated the function of this beam splitter, while experimental proof of its coherent nature remains to be an interesting task for future work. Furthermore, in the future, our general method may be extended or combined with other methods. One could, for example, implement two or more such local beam splitters for increased interferometric control of chemical reactions. In addition, it is very promising to combine our scheme acting on the exit channels, with a scheme acting on the entrance channel (such as the Feshbach scheme in \cite{Haze2025}) for tailoring product distributions on an unprecedented level. 
 
\paragraph{Acknowledgements}
This work was financed by the Baden-W\"{u}rttemberg Stiftung through the Internationale Spitzenforschung program (contract BWST ISF2017-061) and by the German Research Foundation (DFG, Deutsche Forschungsgemeinschaft) within contract 399903135. J.H.D and J.P.D. acknowledge funding by Q-DYNAMO (EU HORIZON-MSCA-2022- SE-01) within project No. 101131418.
We acknowledge support from bwFor-Cluster JUSTUS 2 for high performance computing. J. P. D. also acknowledges partial support from the U.S. National Science Foundation, Grant No. PHY-2012125 and PHY-2308791, and NASA/JPL 1502690. S.H. also acknowledges support from Japan Science and Technology Agency Moonshot R\&D Grant No. JPMJMS2063, ASPIRE Grant No. JPMJAP2319 and PRESTO Grant No. JPMJPR2459.

%\bibliography{refs}

\begin{thebibliography}{45}
\bibitem [{Res()}]{Rescaling}%
	\BibitemOpen
	\href@noop {} {\bibinfo {title} {The data for $({J}' = 5)$ stem from a
			separate data set measured at slightly different {Rb} cloud densities,
			resulting in ion rates and loss signals being globally different by a few
			percent. for an accurate comparison of the different data sets, the $({J}' =
			5)$ data were scaled by the corresponding factor}}\BibitemShut {NoStop}%
\bibitem [{Bfi()}]{Bfield}%
	\BibitemOpen
	\href@noop {} {\bibinfo {title} {Magnetic fields are determined with a
			relative uncertainty of 1.5\%}}\BibitemShut {NoStop}%
\end{thebibliography}


\newpage

\author{Dominik Dorer}
\affiliation{Institut f\"{u}r Quantenmaterie and Center for Integrated Quantum Science \\ and Technology IQ$^{ST}$, Universit\"{a}t Ulm, D-89069 Ulm, Germany}

\author{Shinsuke Haze}
\affiliation{Institut f\"{u}r Quantenmaterie and Center for Integrated Quantum Science \\ and Technology IQ$^{ST}$, Universit\"{a}t Ulm, D-89069 Ulm, Germany}
\affiliation{Center for Quantum Information and Quantum Biology, Osaka University, 1-2 Machikaneyama, Toyonaka, Osaka 560-0043, Japan}

\author{Jing-Lun Li}
\affiliation{Institut f\"{u}r Quantenmaterie and Center for Integrated Quantum Science \\ and Technology IQ$^{ST}$, Universit\"{a}t Ulm, D-89069 Ulm, Germany}

\author{Jos\'{e} P. D'Incao}
\affiliation{Institut f\"{u}r Quantenmaterie and Center for Integrated Quantum Science \\ and Technology IQ$^{ST}$, Universit\"{a}t Ulm, D-89069 Ulm, Germany}
\affiliation{JILA, NIST and Department of Physics, University of Colorado, Boulder, CO 80309-0440, USA}
\affiliation{Department of Physics, University of Massachusetts Boston, Boston, MA 02125, USA }

\author{Eberhard Tiemann}
\affiliation{Institut f\"ur Quantenoptik, Leibniz Universit\"at Hannover, 30167 Hannover, Germany}

\author{Paul S. Julienne}
\affiliation{Institut f\"{u}r Quantenmaterie and Center for Integrated Quantum Science \\ and Technology IQ$^{ST}$, Universit\"{a}t Ulm, D-89069 Ulm, Germany}
\affiliation{Joint Quantum Institute, University of Maryland and NIST, College Park, MD 20742, USA}
\author{Markus Dei{\ss}}
\affiliation{Institut f\"{u}r Quantenmaterie and Center for Integrated Quantum Science \\ and Technology IQ$^{ST}$, Universit\"{a}t Ulm, D-89069 Ulm, Germany}

\author{Johannes Hecker Denschlag} \email{johannes.denschlag@uni-ulm.de}
\affiliation{Institut f\"{u}r Quantenmaterie and Center for Integrated Quantum Science \\ and Technology IQ$^{ST}$, Universit\"{a}t Ulm, D-89069 Ulm, Germany}

\begin{center}
	\Large{\textit{Supplemental Material}}
\end{center}\vspace{-3mm}
\maketitle

\section{Three-body numerical simulation}
\begin{figure}[h]
	\renewcommand{\thefigure}{S1}
	\includegraphics[width=\columnwidth]{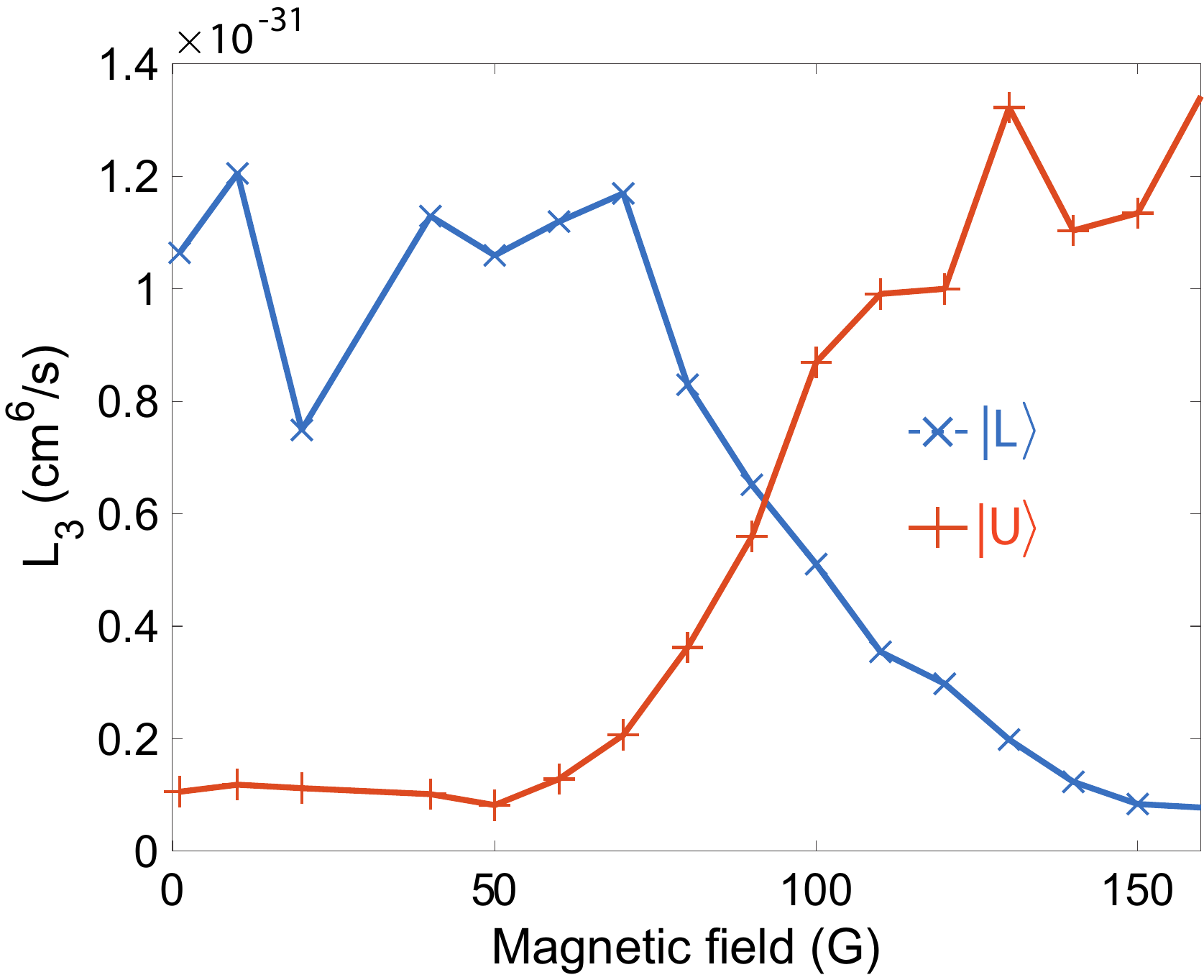}
	\caption{Calculated three-body recombination rate coefficients. The orange (blue) line corresponds to the upper (lower) state.}
	\label{figS1}
\end{figure}

The theoretical recombination partial rate coefficients $L_3$ in Fig. \ref{figS1} are obtained by numerically solving the three-body Schr\"{o}dinger equation in a hyperspherical coordinate representation \cite{Suno2002,wang2011pra}. To properly describe the essential molecular spin mixing between the avoided crossing levels and its role in three-body recombination, our model contains the exact atomic spin structure as in our previous works \cite{Haze2025,Li2025}. The interatomic interactions are taken as the pairwise singlet and triplet Born-Oppenheimer (BO) potentials from \cite{Strauss2010} with, however, the depth being restricted by adding a term of $C_6 \lambda^6/r^{12}$. Here $C_6$ is the van der Waals dispersion coefficient and $r$ denotes the internuclear distance. We tune the short-range parameter $\lambda$ to limit the number of $s$-wave bound states to 6 for both BO potentials and reproduce the low-energy scattering and bound-state properties of the original potentials \cite{Li2025b}. The tuning procedure requires a slight shift of atomic hyperfine coupling as well. The construction of the shallower singlet and triplet potentials and their implementation in the three-body model in hyperspherical coordinate representation have been detailed in Ref. \cite{Li2025b} and Refs. \cite{Haze2025,Li2025}, respectively. To briefly summarize, we use $\lambda = 25.89676 a_0$ and $\lambda = 25.95295 a_0$ for singlet and triplet potential respectively, while the atomic hyperfine coupling strength is reduced by a factor of 0.9026 \cite{Li2025b}. The rest of the atomic and interaction parameters of the system shall take the realistic physical values. In our simulation, we restrict the spin state of the third atom in its initial state ($|f=1,m_f=-1\rangle$) when the other two form a molecule. This has been proved to be a very good approximation for Rb atoms in previous works \cite{Haze2022,Haze2023,Haze2025,Li2025}.
Comparison of the numerical calculations in Fig. \ref{figS1} with the experimentally observed signals in Fig. 4 of the main text shows a qualitatively good agreement.
The relative variation of signal strength is more pronounced in the experiment than predicted by the theory. For example, whereas the rate increases by a factor of 19 $\pm$ 3 for the upper state in the experiment, the increase is only a factor of 11 in the three-body simulations. Furthermore, the position where the production rates cross is shifted to lower magnetic fields compared to the experimental observation. 
This could be a consequence of using potentials which are restricted in  depth.

\section{Transition matrix elements from $| L \rangle$ and
	$| U \rangle$ towards the intermediate state 
	$| e_i \rangle $}
Our calculations show that to a very good approximation, the magnetic field $\mathcal{B}$ dependent states  $| L (\mathcal{B}) \rangle$ and $| U (\mathcal{B}) \rangle$  can be expressed as a superposition  of the molecular states  $|  L (0) \rangle$ and $| U (0) \rangle$  within the magnetic field range considered in our work,
\begin{align}
	| L (\mathcal{B}) \rangle & = \cos (\alpha) \,  | L (0)   \rangle
	-\sin ( \alpha )  \,  | U (0) \rangle  \label{eq:L}\\
	| U (\mathcal{B}) \rangle & = 
	\sin( \alpha )  \,  | L ( 0 ) \rangle
	+ \cos ( \alpha ) \,  | U (0) \rangle\,,
	\label{eq:U}
\end{align}
where $\alpha$ is a magnetic field dependent mixing angle.

Let the electric dipole transition matrix elements towards the excited, intermediate state $ | e_i \rangle $ at zero magnetic field be given as 
\begin{align}
	t_{L,\sigma } (0) &=  \langle e_i  | T_\sigma | L (0) \rangle \\ 
	t_{U,\sigma } (0) &=  \langle e_i  | T_\sigma | U (0) \rangle\,,
\end{align}
where $ T_\sigma$ is the dipole operator and its index 
$\sigma = -1, 0, 1$  indicates the polarization of the excitation light. 
Then the electric dipole transition matrix elements for $| L (\mathcal{B}) \rangle $
and $| U (\mathcal{B}) \rangle $ are
\begin{align}
	t_{L,\sigma} (\mathcal{B}) & =  \cos (\alpha) t_{L,\sigma} (0)   -\sin ( \alpha ) t_{U,\sigma} (0) \\
	t_{U,\sigma} (\mathcal{B}) & =  \sin (\alpha) t_{L,\sigma} (0) +  \cos ( \alpha ) t_{U,\sigma} (0)\,,
\end{align}
which shows that $t_{L,\sigma} (\mathcal{B})$ and $t_{U,\sigma} (\mathcal{B})$ vary with the magnetic field $\mathcal{B}$ due to quantum interference and can even vanish.

The excited intermediate state
$$ | e_i \rangle = | (2)^1\Sigma_u^+ , \mbox{v}' = 36, J' = 5, I' = 2 \rangle $$
has a hyperfine substructure according to the nuclear spin $I' = 2$.
It consists of a group of nearly degenerate levels with 
quantum numbers $F'_{tot} = 7, 6, 5, 4, 3 $ and the corresponding  $ m'_{F_{tot}}$ levels. Here, $F'_{tot}$ is the total angular momentum quantum number ($\vec{F'_{tot}} = \vec{J'} + \vec{I'}$). Therefore, to completely describe one of these excited levels we introduce the notation
$| e_i, F'_{tot}, m'_{F_{tot}} \rangle$.
These states are essentially 
$\mathcal{B}$-field independent, as a consequence of the $^1\Sigma$ electronic state. 
Similarly, also the  levels  $| L (0) \rangle $
and $| U (0) \rangle $ have a substructure in terms of the quantum number $m_{L_R} = -6, \dots, 6$ for $\mathcal{B} > 0$ (and
in terms of $F_{tot} = 4, \dots,6$ for
$\mathcal{B} = 0$). Hence, we label such individual quantum states by  
$| L(0),  m_{L_R}  \rangle$ and
$| U(0),  m_{L_R}  \rangle$. In the experiment, the
$m_{L_R}$ levels are uniformly populated,
(but this is not resolved experimentally as the corresponding level splittings are only on the order of MHz). 

Coupled channel calculations show an almost constant ratio of 
$ t_{L,\sigma} (0) / t_{U,\sigma} (0) = 0.576 $ (variation less than one percent), independent of the sublevels and of the polarization of the light. This independence results directly from the extremely weak coupling of $\vec{F}$ with $\vec{L}_R$ to   $\vec{F}_{tot}$ which has a magnitude on the order of one MHz$\times h$. In comparison, the hyperfine splittings
between different $F$ levels are in the GHz$\times h$ range.

\begin{figure}[h]
	\renewcommand{\thefigure}{S2}
	\includegraphics[width=0.87\columnwidth]{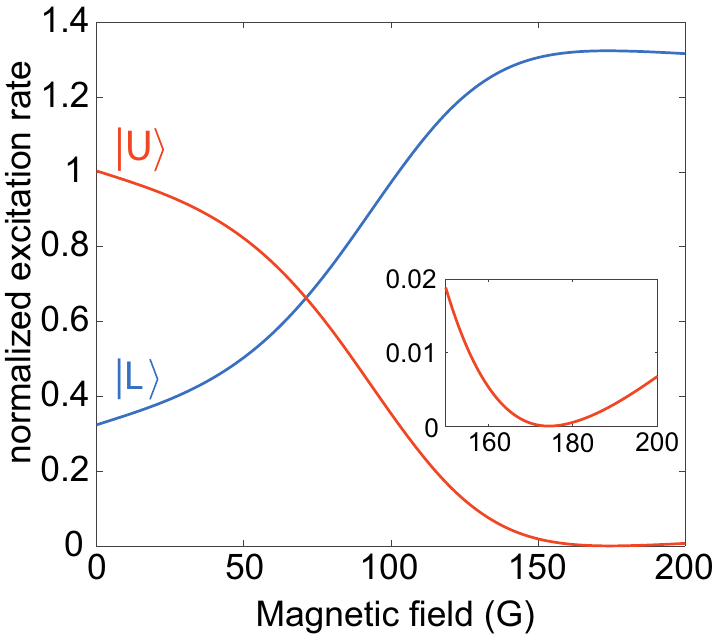}
	\caption{Normalized transition rates from the states
		$| L (\mathcal{B}) \rangle $ and $| U (\mathcal{B})  \rangle $ towards the intermediate state $| e_i \rangle$ in the non-saturated regime. The inset is a blowup of the high-field section. 
	}
	\label{figS2}
\end{figure}

Figure \ref{figS2} shows the calculated excitation rate (in the limit of no saturation) of the 
levels $| U (\mathcal{B}) \rangle $  and $| L (\mathcal{B}) \rangle $ towards the intermediate state as a function of magnetic field $\mathcal{B}$.
The rates are normalized to the excitation rate for $| U (\mathcal{B}) \rangle $ at $\mathcal{B} = 0$.  The curves are the same for the polarizations $\pi, \sigma_{\pm}$.
The calculations predict that the excitation rate for $| U (\mathcal{B}) \rangle $ vanishes at $\mathcal{B} \approx $175 G (see inset of Figure \ref{figS2}). For higher fields it rises again. 

\begin{figure}[h]
	\renewcommand{\thefigure}{S3}
	\includegraphics[width=\columnwidth]{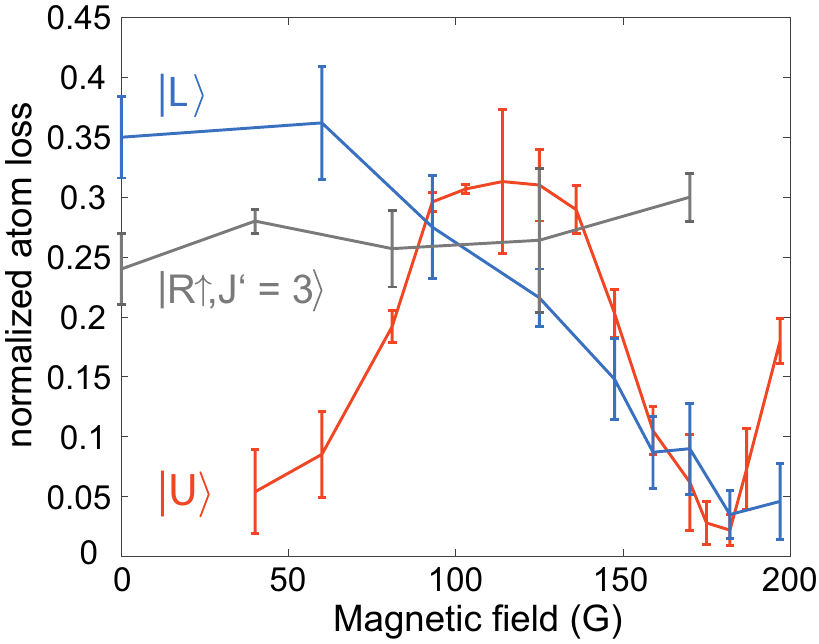}
	\caption{
		REMPI signals of molecules which have been produced via TBR. Here, the first REMPI transition was generally not saturated. The measurements were carried out using the simple, qualitative scheme that was employed for Fig. 3 of the main part of the publication. The REMPI laser frequency was set on the resonance position for each magnetic field. We show the normalized atom loss which roughly represents the ion production rate. The ion production rate, in turn, is roughly proportional to the product of 
		the molecular production rate due to TBR and the optical excitation rate towards the intermediate state. 
		Each data point represents the mean value of four repetitions of the experiment and the error bars correspond to one standard deviation.}
	\label{figS3}
\end{figure}

In order to experimentally test the strong variation of the excitation rate 
for $| U (\mathcal{B}) \rangle $ and in particular its vanishing at around 175 G, we carried out spectroscopy with low laser power, so that the excitation rate toward the intermediate state $|e_i \rangle$ is generally not saturated. The experiments were carried out following the simple, qualitative scheme that we also used for the data in Fig. 3 in the main text. 
For each magnetic field, the REMPI laser was
set on the resonance position and 
the atomic loss was measured after a fixed time. 
Figure \ref{figS3} shows the results.
The atom loss is roughly proportional to the ion production rate which, in turn, is roughly proportional to the product of the molecular production rate and the optical excitation rate towards the intermediate level.
At low $\mathcal{B}$-field, the signals for $| U (\mathcal{B}) \rangle $ 
are below the detection limit because the three-body recombination flux into that channel is too small. With increasing $\mathcal{B}$-field the signal at first increases because of the increasing reaction rate into channel $| U (\mathcal{B}) \rangle $. 
Beyond $\mathcal{B} = 120$ G the signal decreases again and gets minimal 
at around 175 G, due to the vanishing of the excitation rate (as predicted in Fig. \ref{figS2}). For larger $\mathcal{B}$-fields the signals reappear.
The signals for $| L (\mathcal{B}) \rangle $ decrease with the $\mathcal{B}$-field because the three-body recombination rate decreases, despite a factor of four increase in the  excitation rate. The signal for the reference level stays relatively constant, as expected. Therefore, the experimental results confirm our theoretical expectations.

\end{document}